\documentclass[3p,times,twocolumn]{elsarticle}

\usepackage{ecrc}

\volume{00}

\firstpage{1}

\journalname{Nuclear Physics B Proceedings Supplement}

\runauth{L.~Hofer et al.}

\jid{nuphbp}

\jnltitlelogo{Nuclear Physics B Proceedings Supplement}

\usepackage{amssymb}

\usepackage[figuresright]{rotating}

\begin{document}

\begin{frontmatter}



\dochead{}

\title{QCD uncertainties in the prediction of $B\to K^*\mu^+\mu^-$ observables}


\author[lab1]{S.~Descotes-Genon}
\author[lab2]{L.~Hofer\corref{cor1}}
\author[lab2]{J.~Matias}
\author[lab3]{J.~Virto}
\cortext[cor1]{Talk given by L.~Hofer at the 
37$^{th}$ International Conference on High Energy Physics (ICHEP 2014).}

\address[lab1]{Laboratoire de Physique Th\'eorique, CNRS/Univ. Paris-Sud 11 (UMR 8627), 91405 Orsay Cedex, France}
\address[lab2]{Universitat Aut\`onoma de Barcelona, 08193 Bellaterra, Barcelona, Spain}
\address[lab3]{Theoretische Physik 1, Naturwissenschaftlich-Technische Fakult\"at,
Universit\"at Siegen, 57068 Siegen, Germany}

\begin{abstract}
The recent LHCb angular analysis of the exclusive decay $B\to K^* \mu^+ \mu^-$ 
has indicated significant deviations from the Standard Model expectations. In order to give precise 
theory predictions, it is crucial that uncertainties from non-perturbative QCD are under control and 
properly included. The dominant QCD uncertainties originate from the hadronic 
$B\to K^*$ form factors and from $c\bar{c}$ loops. We present a systematic method to include 
factorisable power corrections to the form factors in the framework of QCD factorisation and study the impact of 
the scheme chosen to define the soft form factors. We also discuss charm-loop effects.
\end{abstract}

\begin{keyword}
rare B decays \sep LHCb phenomenology \sep QCD factorisation

\end{keyword}

\end{frontmatter}


\section{Introduction}
\label{sec:intro}

The semi-leptonic decay $B\to K^*\mu^+\mu^-$ with the vector-meson $K^*$ subsequently decaying as 
$K^*\to K\pi$ constitutes an ideal channel for the search for new physics (NP) beyond the standard model (SM). 
An angular analysis of the full four-body final-state allows to construct sets of observables whose 
experimental measurement is not only
able to reveal a possible deviation from the SM prediction but is even capable of discriminating 
different models of NP. As a drawback, the extraction of information on high-scale new physics is hampered
by the impact of non-perturbative QCD effects, entering mainly through hadronic $B\to K^*$ form factors and 
resonant $c\bar{c}$ intermediate states. 
The sensitivity to form factors can be significantly reduced 
by considering appropriate observables and an optimised set of such observables 
has been given in ref.~\cite{Descotes-Genon:2013vna} for the region of
large hadronic recoil, i.e. for a small invariant mass $q^2\lesssim 8\,\rm{GeV}^2$ 
of the muon pair.\pagebreak
 
In this region of large hadronic recoil,
the recent LHCb angular analysis \cite{Aaij:2013iag,Aaij:2013qta} 
has indicated significant deviations from SM expectations, most notably in the observables 
$P_5^\prime$ \cite{DescotesGenon:2012zf} 
and $P_2$ \cite{Becirevic:2011bp,Matias:2012xw}. 
The upcoming analysis with an increased amount of data will show if these deviations are physical
effects or only statistical fluctuations. To this end relations among the observables can be used 
to check consistence of the experimental results  
\cite{nicola}. If the anomaly persists, it can 
be accomodated in NP scenarios with an additional contribution to the Wilson coefficient $C_9$ of
about $-25\%$ of its SM value, as pointed out originally in 
ref.~\cite{Descotes-Genon:2013wba}. This basic observation has been confirmed by independent studies
using a different set of observables and/or statistical 
methods~\cite{Altmannshofer:2013foa,Beaujean:2013soa,Horgan:2013pva,Hurth:2014vma}.

In order to be able to draw solid conclusions on potential high-scale NP effects from $B\to K^*\mu^+\mu^-$ data, 
it is important that uncertainties from non-perturbative QCD are under control and properly included in
the theory predictions. In this proceeding we discuss the dominant uncertainties stemming from the hadronic 
$B\to K^*$ form factors and from $c\bar{c}$ loops, summarising our results from ref.~\cite{Descotes-Genon:2014uoa}. 

\section{Factorisable power corrections}
\label{sec:FF}

\subsection{Soft form factors}

The evaluation of matrix elements for the decay $B\to K^*\mu^+\mu^-$ involves seven non-perturbative form factors
$V,A_{0,1,2},T_{1,2,3}$ (see ref.~\cite{Beneke:2000wa} for definitions). LCSR calculations of these form factors
\cite{Ball:2004ye,Khodjamirian:2010vf} suffer from large uncertainties originating from hadronic parameters, 
and moreover rely on certain assumptions 
(modelling the continuum contribution, suppression of excited states, etc.) 
introducing systematic uncertainties that are difficult to quantify\footnote{It is reasonable 
to assume a 10\% irreducible uncertainty from these sources \cite{Braun}.}. 
Furthermore, LCSR results are usually presented without specifying the correlations among the various
form factors.

In the region of large recoil and at leading order in $\alpha_s$ and $\Lambda/m_b$, 
heavy-quark symmetries relate the seven form factors $V,A_{0,1,2},T_{1,2,3}$ among each other, reducing
the number of independent form factors to two~\cite{Beneke:2000wa,Charles:1998dr,Bauer:2000yr}. Different choices are possible for the selection of these
two so-called soft form factors:
\begin{eqnarray}
&\{V,A_0,A_1,A_2,T_1,T_2,T_3\}\nonumber\\
&\bf \Downarrow&\\
&\{V,A_0\}\;\textrm{  or  }\;
    \{V,A_{12}\}\;\textrm{  or  }\;
    \{T_1,A_0\}\;\textrm{  or  }\;...\;.\nonumber&
\end{eqnarray}
Here $A_{12}$ represents the linear combination
\begin{equation}
  A_{12}(q^2)=\frac{m_B+m_{K^*}}{2E}A_1(q^2)-\frac{m_B-m_{K^*}}{m_B}A_2(q^2)
  \label{eq:xiA12}
\end{equation}
of the form factors $A_1$ and $A_2$ with $E$ denoting the energy of the $K^*$ meson, and $m_B$ and $m_{K^*}$
the masses of the $B$- and $K^*$-meson, respectively. By expressing the seven form factors in terms of 
two soft form factors, hadronic uncertainties from the
LCSR input are significantly reduced because dominant correlations are automatically taken into account.

Higher orders in $\alpha_s$ and $\Lambda/m_b$ break the large-recoil symmetry relations among form factors. 
While effects of order $\alpha_s$ can be consistently included in the analysis using the framework of 
QCD factorisation (QCDF)
\cite{Beneke:2000wa,Beneke:2001at,Grinstein:2004vb}, effects of order $\Lambda/m_b$ can only be 
estimated\footnote{Higher-order effects 
are in principle fully included in the LCSR results for the form factors. However, using these
results requires knowledge of their correlations to at least the same precision as they can be infered from
large-recoil symmetries, and leads to results with a stronger dependence on the LCSR input}.

\begin{figure}[t]
\includegraphics[width=7.5cm]{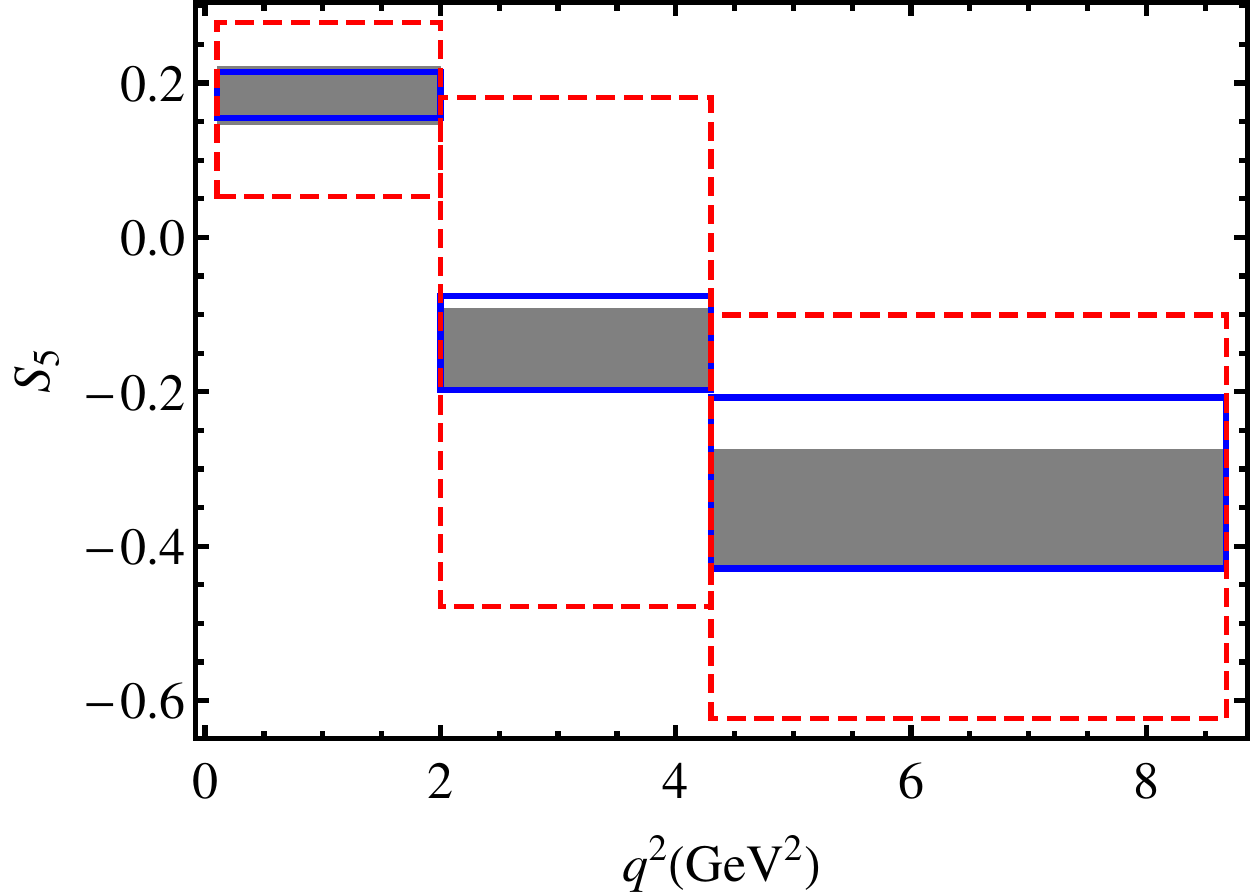}\vspace{7mm}
\includegraphics[width=7.5cm]{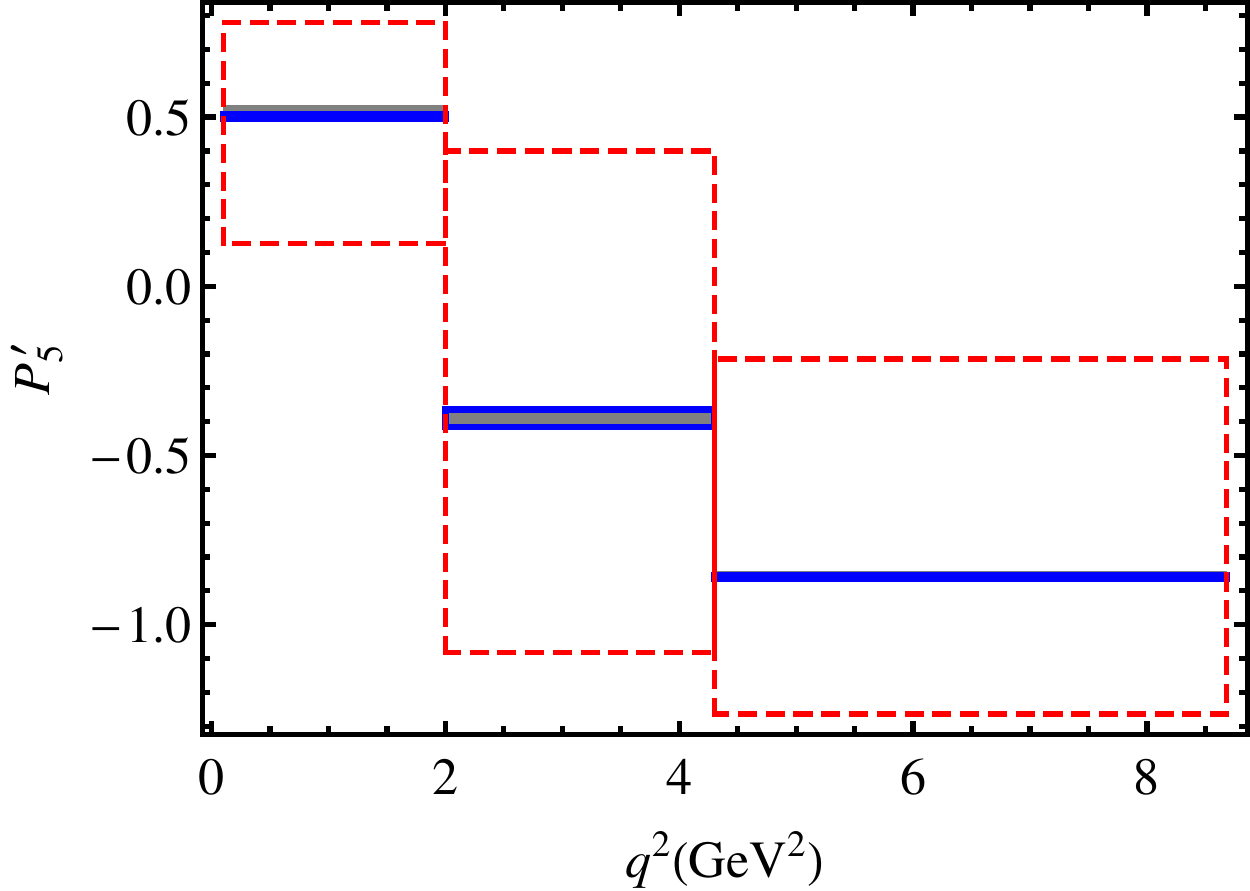}
\caption{Scheme dependence in the prediction of the observables $S_5$ and $P_5'$ at LO in $\Lambda/m_b$. 
Uncertainties are solely due to form factors. Grey bands correspond to the renormalisation scheme 
with LCSR input $\{V,A_{12}\}$ and blue (solid) boxes to the
scheme with input $\{T_1,A_0\}$. Red (dashed) boxes display results obtained using
the full set of form factors \emph{without correlations}. Form factor input
is taken from ref.~\cite{Khodjamirian:2010vf} in all cases.}
\label{fig:LO}
\end{figure}

The choice of the two soft form factors
defines a renormalisation scheme, and theory predictions made to a certain order in $\alpha_s$ or
$\Lambda/m_b$ will exhibit a scheme dependence at the level of the neglected higher orders. This scheme dependence
is illustrated in fig.~\ref{fig:LO} at order $\mathcal{O}(\alpha_s)$ but at leading order in $\Lambda/m_b$ for
the observable $S_5$~\cite{Altmannshofer:2008dz}
and the optimised observable $P_5^{\prime}$~\cite{DescotesGenon:2012zf} in two different schemes 
(grey bands and blue(solid) boxes). The optimsed
observables are constructed in such a way that any dependence on form factors drops out at LO in 
$\alpha_s$ and $\Lambda/m_b$. Therefore the uncertainty associated to the form factor input as well as the scheme
dependence are pushed to order $\mathcal{O}(\alpha_s)$ for the observable $P_5^\prime$ in contrast 
to the observable $S_5$. In addition we show the prediction which one would obtain using uncorrelated
QCD form factors without resorting to large-recoil symmetries (red (dashed) boxes). The result 
demonstrates that in absence of a precise knowledge of correlations it is indispensable to make use of
the soft form factor decomposition.

\subsection{Including power corrections}

Since form factor uncertainties enter optimised observables only at order $\mathcal{O}(\alpha_s,\Lambda/m_b)$,
$\Lambda/m_b$ corrections to large-recoil symmetry relations, called factorisable power corrections, are expected
to be of the same order of magnitude. It is thus desirable to include them into the soft form factor decomposition.
Even though there does not exist a direct calculation of these corrections, they can be assessed indirectly as they
are contained in the LCSR results for the full form factors. Starting from a parametrisation
\begin{equation}
F^{\rm LCSR}(q^2)=F^{\rm soft}(q^2)+\Delta F^{\alpha_s}(q^2)+a_F+b_Fq^2+... 
\label{fig:FFpara}
\end{equation}
of the full LCSR form factors $F^{\rm LCSR}$, with $F^{\rm soft}$ representing the LO expression in the
large-recoil limit and $\Delta F^{\alpha_s}$ the QCDF corrections, 
information on factorisable power corrections encoded in the parameters $a_F,b_F,...$ can be obtained 
from a fit to the full $F^{\rm LCSR}(q^2)$.

This strategy has been proposed and followed for the first time in ref.~\cite{Jager:2012uw}. The authors
of ref.~\cite{Jager:2012uw} fit the parameters $a_F,b_F$ using central values for the form factors
$F^{\rm LCSR},F^{\rm soft}$, and they interpret the result $\hat{a}_F,\hat{b}_F$ as an order of magnitude 
estimate for the power corrections. In this spirit the uncertainties associated to
power corrections are estimated by varying independently $-|\hat{a}_F|<a_F<|\hat{a}_F|$ and 
$-|\hat{b}_F|<b_F<|\hat{b}_F|$. In this approach, the central values of theory predictions are not affected
by power corrections, in particular their scheme dependence is not reduced. Furthermore, the uncertainties
associated with the power corrections are determined from the central values of the form factors, 
even though from the conceptual 
point of view they are related to the uncertainties of the latter. In particular in the hypothetical case
in which the form factors are precisely known, i.e. their uncertainties (and the
uncertainties of their power corrections) go to zero, the obtained error estimate for the power corrections
would remain constant and different from zero\footnote{The only exception is given by 
the accidental situation in which the power corrections themselves were zero.}.  
Finally, the method does not make use of the full information obtained from the fit as the definite and correlated
signs of the fit parameters $\hat{a}_F,\hat{b}_F$ get lost. In the light of the definite sign of the fit results,
the symmetric variation of the $a_F,b_F,...$ around zero implies that power corrections are underestimated in
one direction while they are overestimated in the other.

In our analysis we modify the approach of ref.~\cite{Jager:2012uw} and go beyond it in several aspects. We keep
the fit values $\hat{a}_F,\hat{b}_F,...$ as non-zero central values and vary 
\begin{equation}
   \hat{a}_F-\Delta \hat{a}_F < a_F < \hat{a}_F+\Delta \hat{a}_F, ...\; .
\label{eq:ErrVar}
\end{equation}
In order to fix the ranges $\Delta \hat{a}_F,\Delta \hat{b}_F,...$ for the variation, we consider an expanded version
$F^{\rm LCSR}(q^2)=A_F+B_Fq^2/m_B^2+...$ of the full LCSR form factors and attribute 
to the power corrections an uncertainty of $10\%$ of the full form factor 
setting $\Delta\hat{a}_F=0.1A_F,\Delta\hat{b}_F=0.1B_F,...\;$. This procedure has the 
following features: The non-zero central values of the power corrections shift the central values of observables
to the values which one would obtain using directly the full form factors. This implies that our predictions
for the central values are scheme-independent\footnote{There is still a small residual scheme dependence at
$\mathcal{O}(\Lambda/m_b)$ introduced by non-factorisable power corrections to the QCDF-amplitude, i.e. 
by power corrections that are not related to the description of the full form factors
in terms of soft form factors.}. The error variation is performed with respect to the shifted central values
implying a shift of the error bands with respect to the error bands obtained in ref.~\cite{Jager:2012uw}. 
Our error estimate is conservative as it amounts to assigning an error of 
$\sim 100\%$ to the result from the fit, given the fact that the typical size of power corrections is
$\Delta F^\Lambda\sim F\times\mathcal{O}(\Lambda/m_b)\sim 0.1F$. Moreover, since the error ranges
$\Delta \hat{a}_F,\Delta\hat{b}_B,...$ are introduced by hand, they can easily be adopted once information
on the uncertainties of the LCSR form factors improves by smaller overall errors or better knowledge of correlations.  

\subsection{Correlations}

Power corrections are constrained, on the one hand from exact kinematic relations to be fulfilled by the
full form factors at $q^2=0$, and on the other hand by the choice of the renormalisation scheme for the 
soft form factors. These correlations, which have to be taken into account when the $a_F,b_F,...$ are varied
within the ranges of eq.~(\ref{eq:ErrVar}), are thus scheme-dependent. Taking for example $\{T_1,A_0\}$ as
soft form factors eliminates power corrections (and the corresponding uncertainties) in the form factors
$T_1$ and $A_0$, while taking $\{V,A_{12}\}$ as input eliminates power corrections in $V$ and minimises their
effects in $A_1,A_2$. A change of the renormalisation scheme corresponds to the reshuffling of power corrections
among the different form factors. An appropriate choice of the renormalisation scheme can therefore 
reduce the impact of power corrections on a certain observable by shifting the power corrections into those
form factors to which the observable is less sensitive. In principle it is possible to choose
for each observable the optimal scheme which minimises its individual error. Note, however, that in a
global analysis one is forced to use the same scheme for all observables if one does not want to loose
correlations among the observables. 

\begin{figure}
\centering
\includegraphics[width=7cm,height=4.6cm]{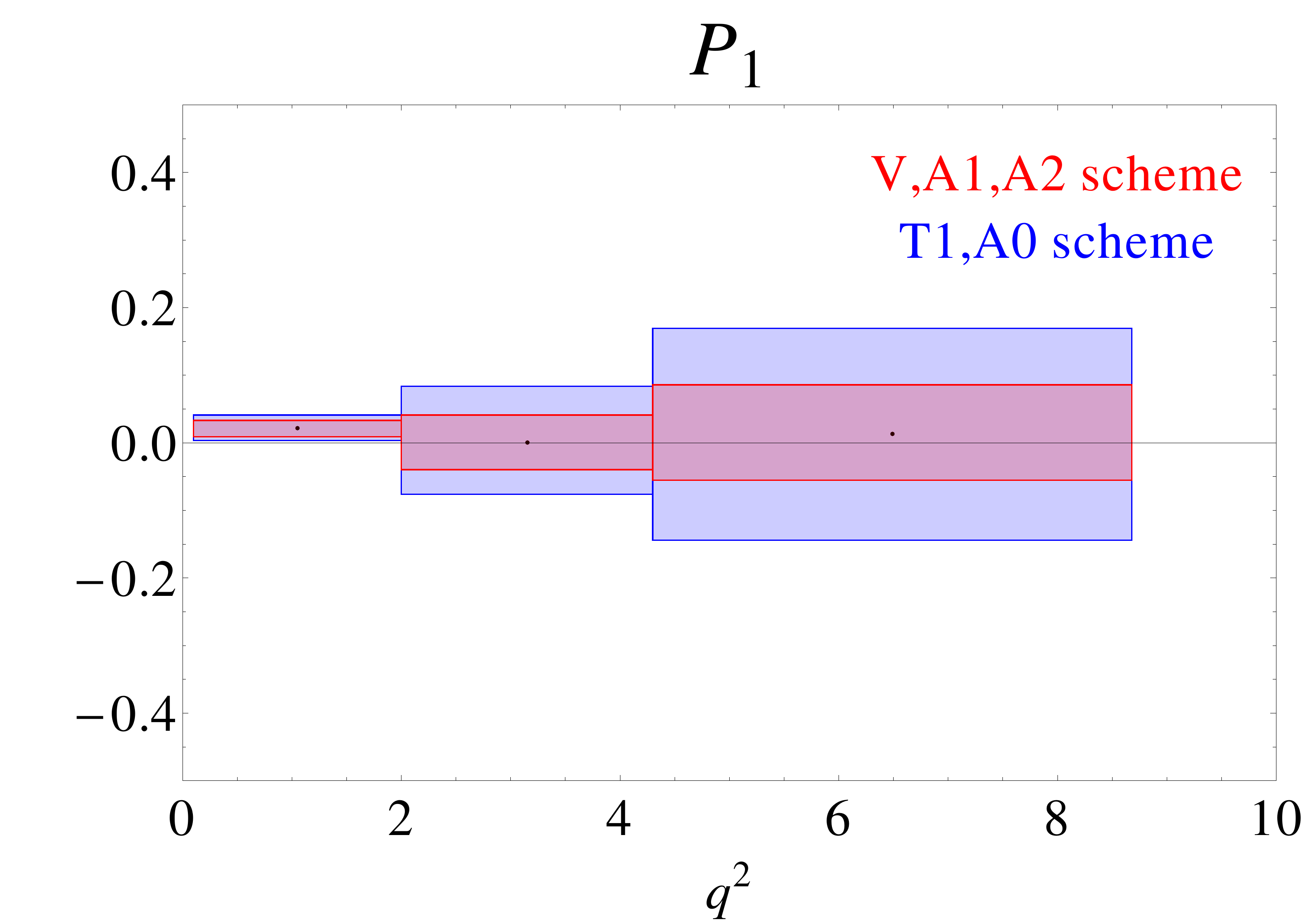}\\[5mm]
\includegraphics[width=7cm,height=4.6cm]{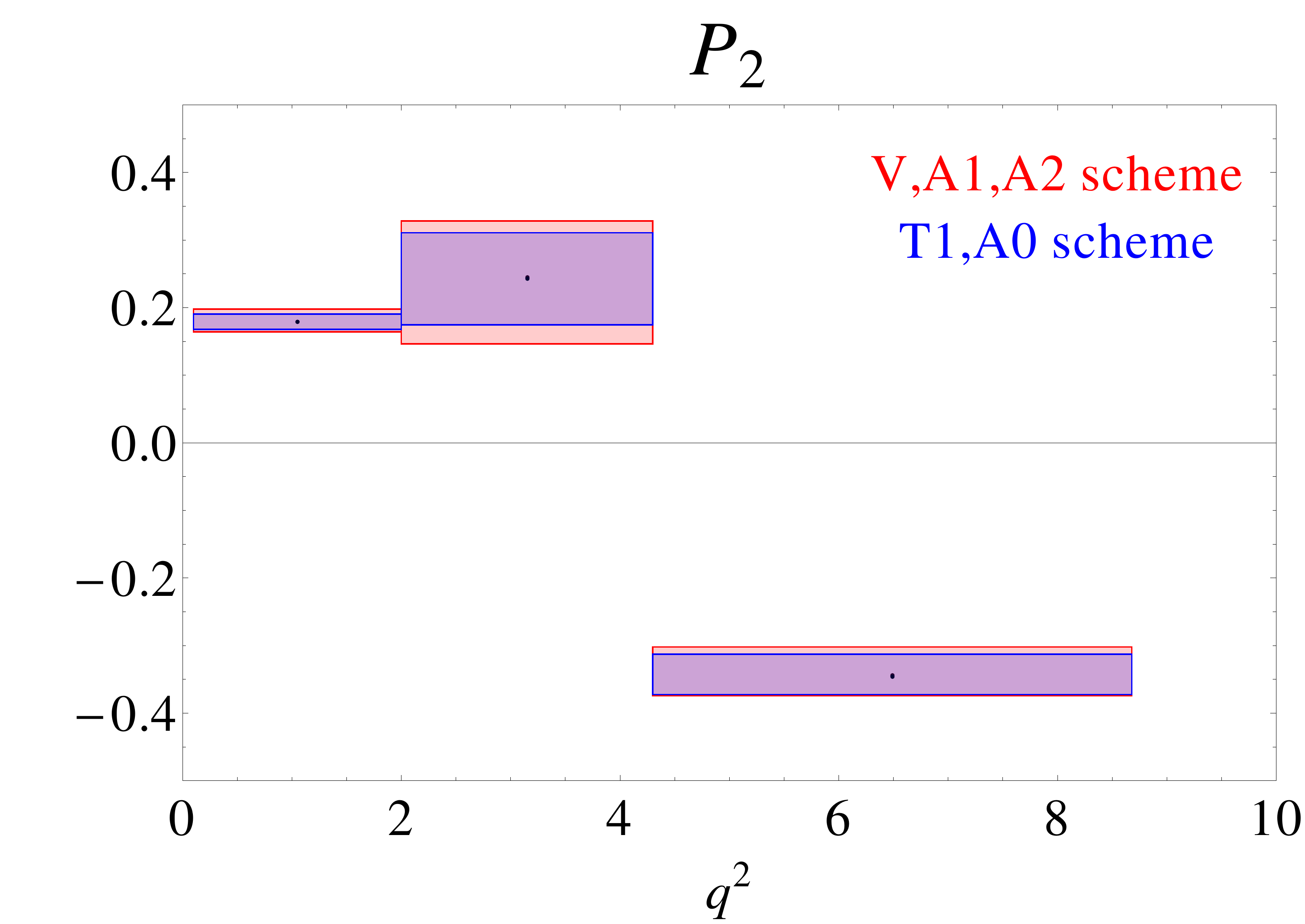}\\[5mm]
\includegraphics[width=7cm,height=4.6cm]{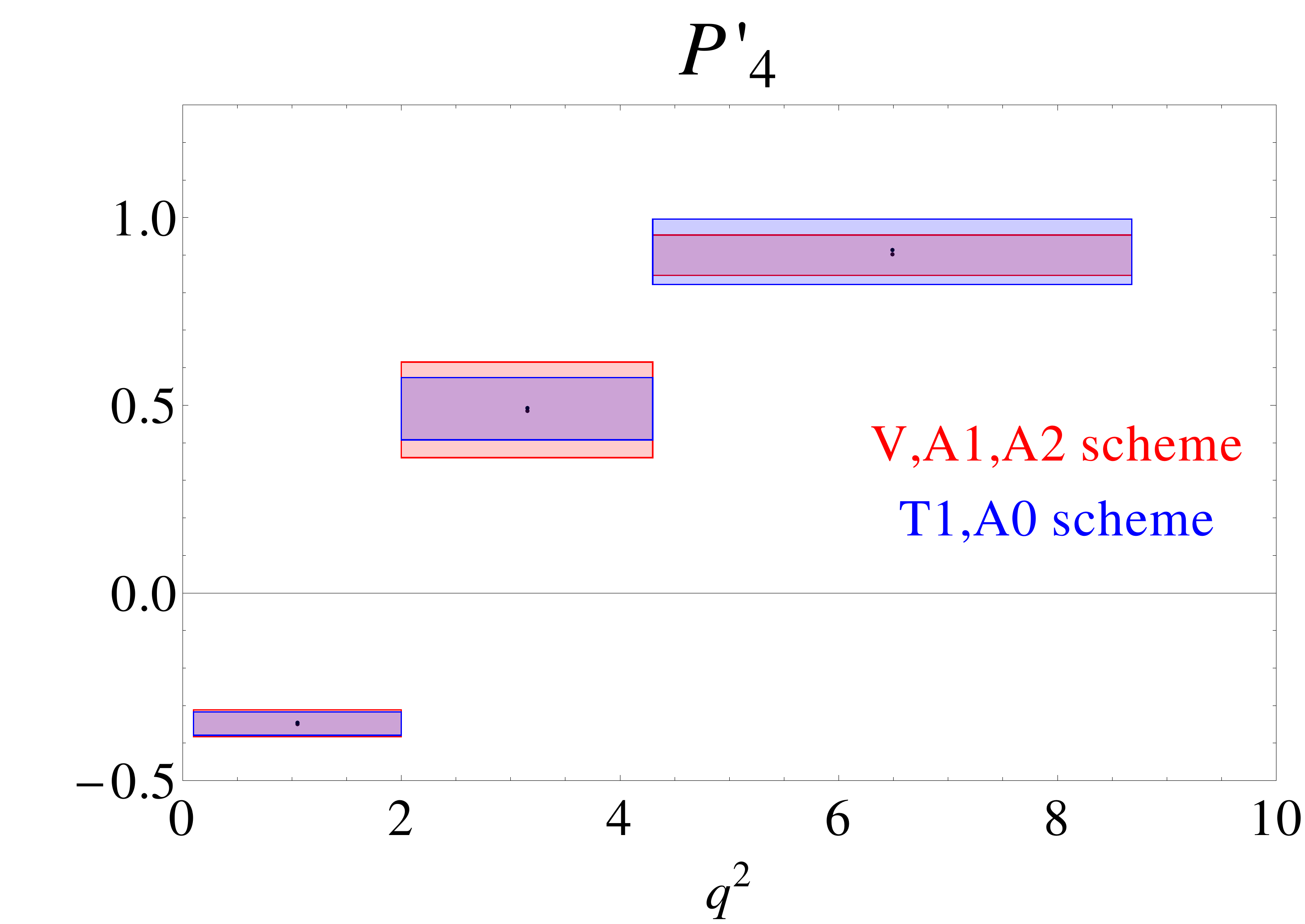}\\[5mm]
\includegraphics[width=7cm,height=4.6cm]{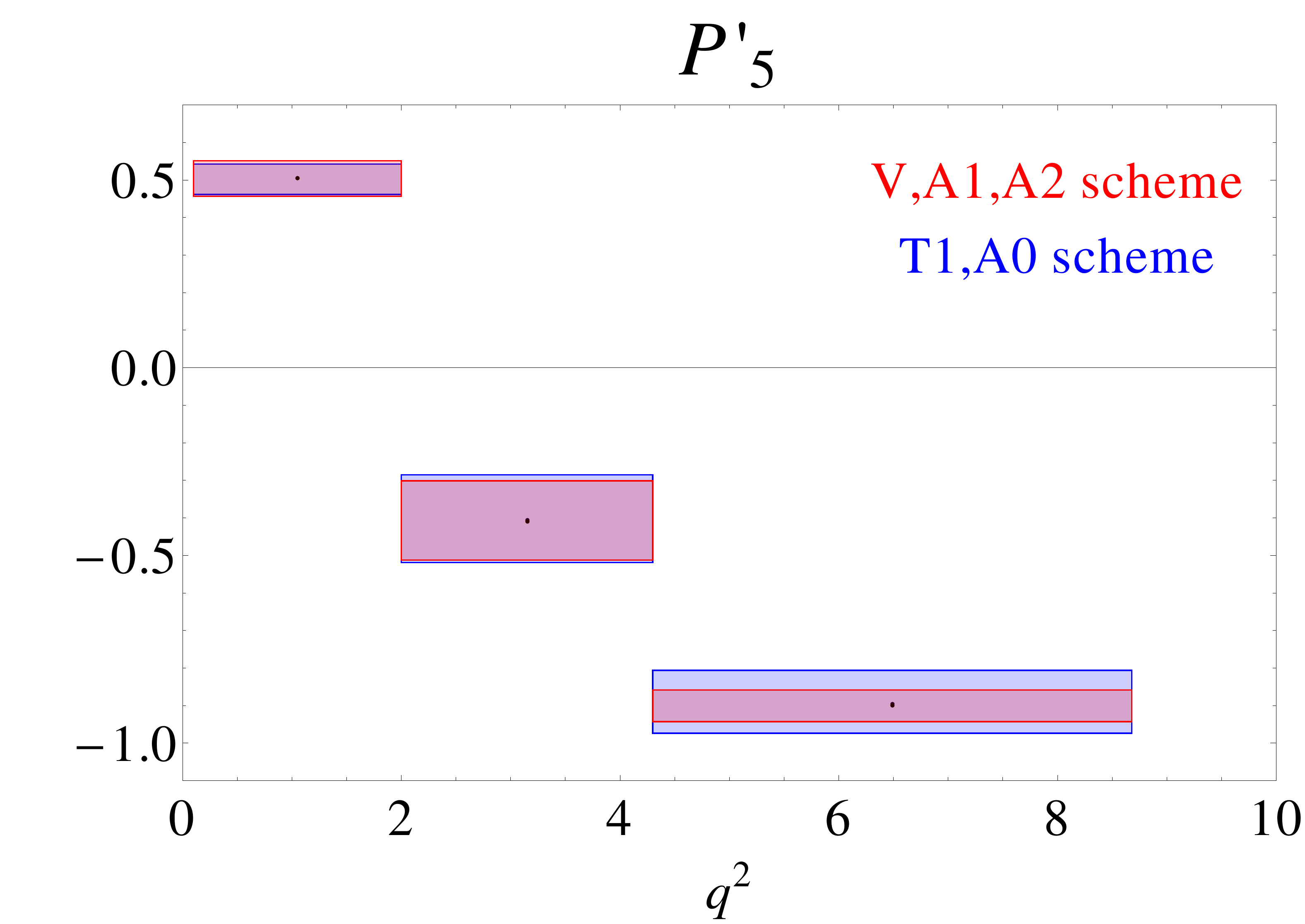}
\caption{Scheme dependence on the prediction of the observables $P_1$, $P_2$, $P_4'$, $P_5'$ in QCD factorisation. 
These results include factorisable power corrections as described in the text.}
\label{fig_SD}
\end{figure}

In fig.~\ref{fig_SD} we show our predictions including power corrections for the observables
$P_1$, $P_2$, $P_4^\prime$ and $P_5^\prime$. We parametrised the power corrections as 
$a_F+b_Fq^2/m_B^2+c_Fq^4/m_B^4$ and we performed a flat scan of the $a_F,b_F,c_F$ over the sub-space allowed
by the correlations. The blue bands represent the results obtained for the renormalisation scheme with
$\{T_1,A_0\}$ as input, while the red bands represent the results for the scheme with $\{V,A_{12}\}$. Since 
the observable $P_5^\prime$ showing the anomaly is much more
sensitive to the vector form factor $V$ than to the tensor form factor $T_1$, and since the contribution
of the form factor $A_0$ to observables is always suppressed by small lepton masses, the $\{V,A_{12}\}$-scheme
is phenomenologically favoured compared to the $\{T_1,A_0\}$-scheme.

In ref.~\cite{Descotes-Genon:2014uoa} we give detailed predictions
in this scheme for all the $S_i$- and $P_i^{(\prime)}$-observables, including apart from the error associated to
the factorisable power corrections also parametric errors, form factor errors and errors corresponding to
non-factorisable power corrections.
We note that for optimised observables and for input taken
from ref.~\cite{Khodjamirian:2010vf}, parametric uncertainties, form factor uncertainties and uncertainties from 
factorisable power corrections are usually of the same order of magnitude, while uncertainties from 
non-factorisable power corrections are typically smaller.
For ''non-optimised observables``, uncertainties are dominated by the form factor input as expected. 
For input taken from ref.~\cite{Ball:2004ye},
the uncertainties stemming from the form factors are generally smaller, in particular they are 
completely negligible for optimised observables.

\section{Long-distance charm loop effects}
\label{sec:cc}

The long-distance contribution 
from $c\bar{c}$ loops does not stand on the same footing as the
factorisable power corrections discussed in the previous section. Its size is a debated issue, 
with some contributions considered 
in ref.~\cite{Khodjamirian:2010vf} for $B\to K^*\mu^+\mu^-$ and further work 
(unfortunately only for $B\to K\mu^+\mu^-$) 
in ref.~\cite{khodjimannel}. Very recently it has been claimed that low-scale non-perturbative and/or
high-scale new physics contributions to the charm loop could explain the anomalous
$B\to K^*\mu^+\mu^-$ data \cite{zwicky}.

For an overall estimate of non-perturbative contributions from hadronic operators, 
we consider the terms $\Delta C_9$ in ref.~\cite{Khodjamirian:2010vf}, 
which include the LO perturbative contribution from $O_{1,2}$ together with non-factorisable soft-gluon emission 
from the charm loop. 
In order to separate the long-distance contribution $\delta C_9^{\rm LD}$,
we subtract the perturbative contribution from  $\Delta C_9$. We add this contribution to each amplitude 
$A_i^{L,R}$ ($i=\perp,\|,0$) by substituting
\begin{equation}
C_9 \to C_9 + s_i \delta C_9^{\rm{LD}}(q^2)\ .
\end{equation}
The result of the calculation in ref.~\cite{Khodjamirian:2010vf} corresponds to setting $s_i=1$. However,
since the computation of ref.~\cite{Khodjamirian:2010vf} does not include all contributions, we prefer to
interpret their result only as an order-of-magnitude estimate and to vary the parameters $s_i$ thus 
in the range $[-1,1]$. Through the independent variation of the $s_i$ we make sure that contributions to 
different amplitudes are not artificially correlated and we allow for the possibility of 
long-distance contributions with opposite signs in the different amplitudes. 

\begin{figure}
\centering
\includegraphics[width=7cm,height=4.6cm]{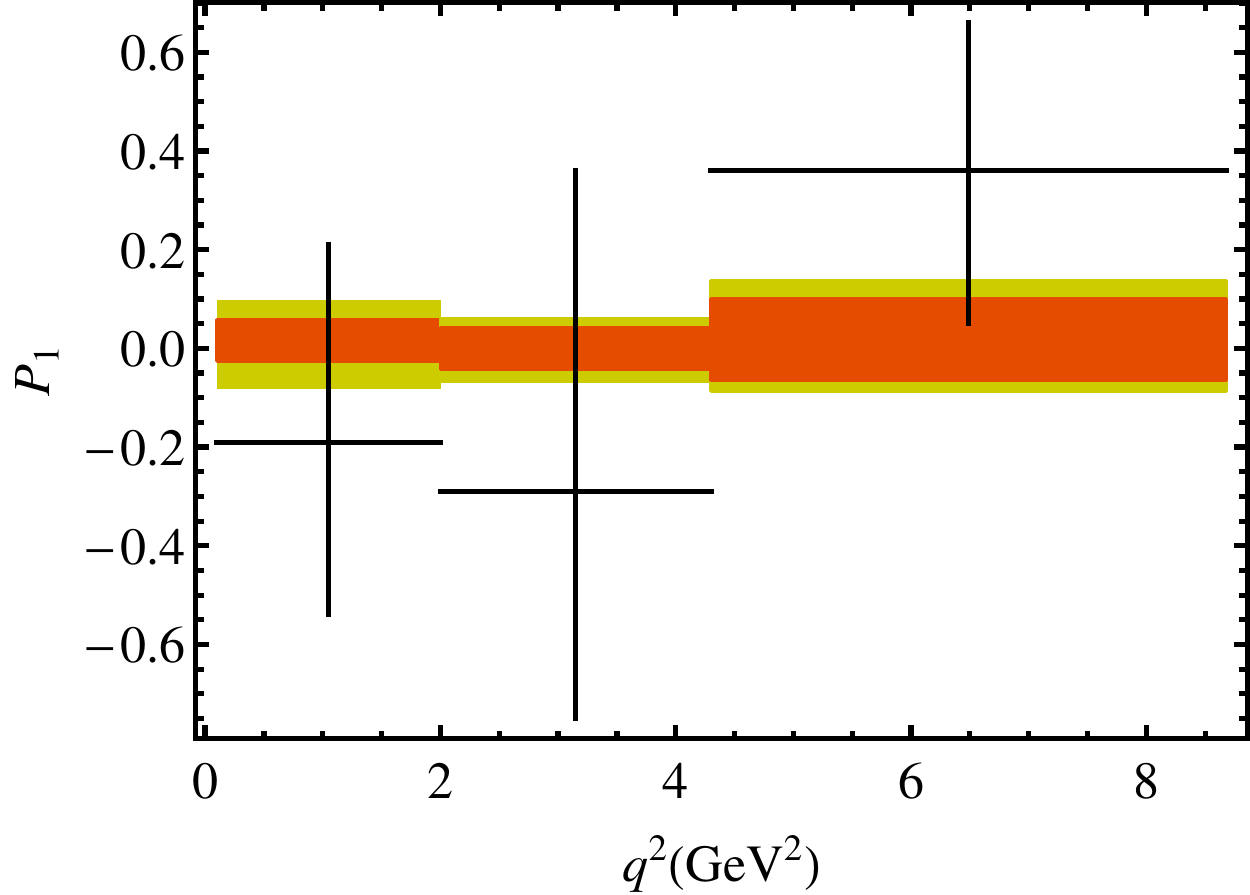}\\[5mm]
\includegraphics[width=7cm,height=4.6cm]{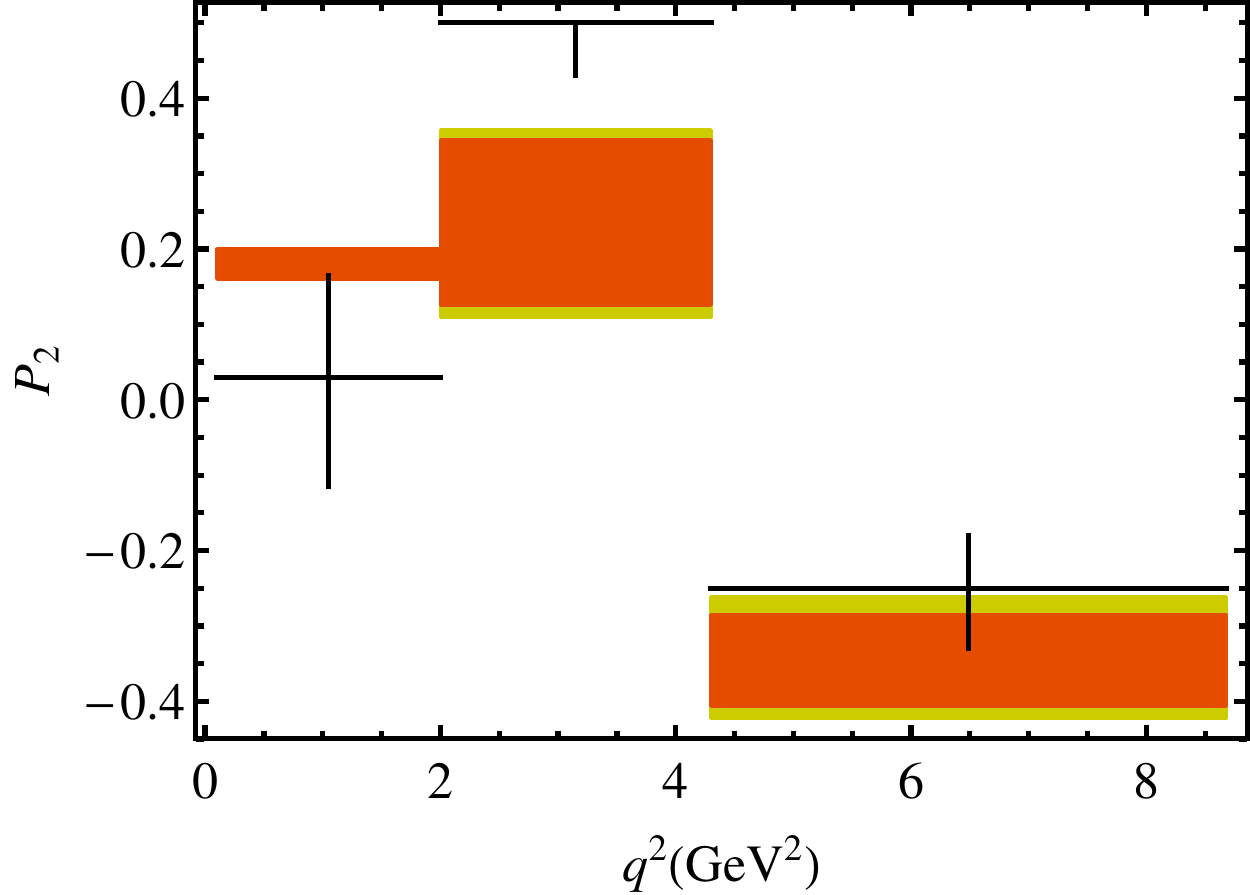}\\[5mm]
\includegraphics[width=7cm,height=4.6cm]{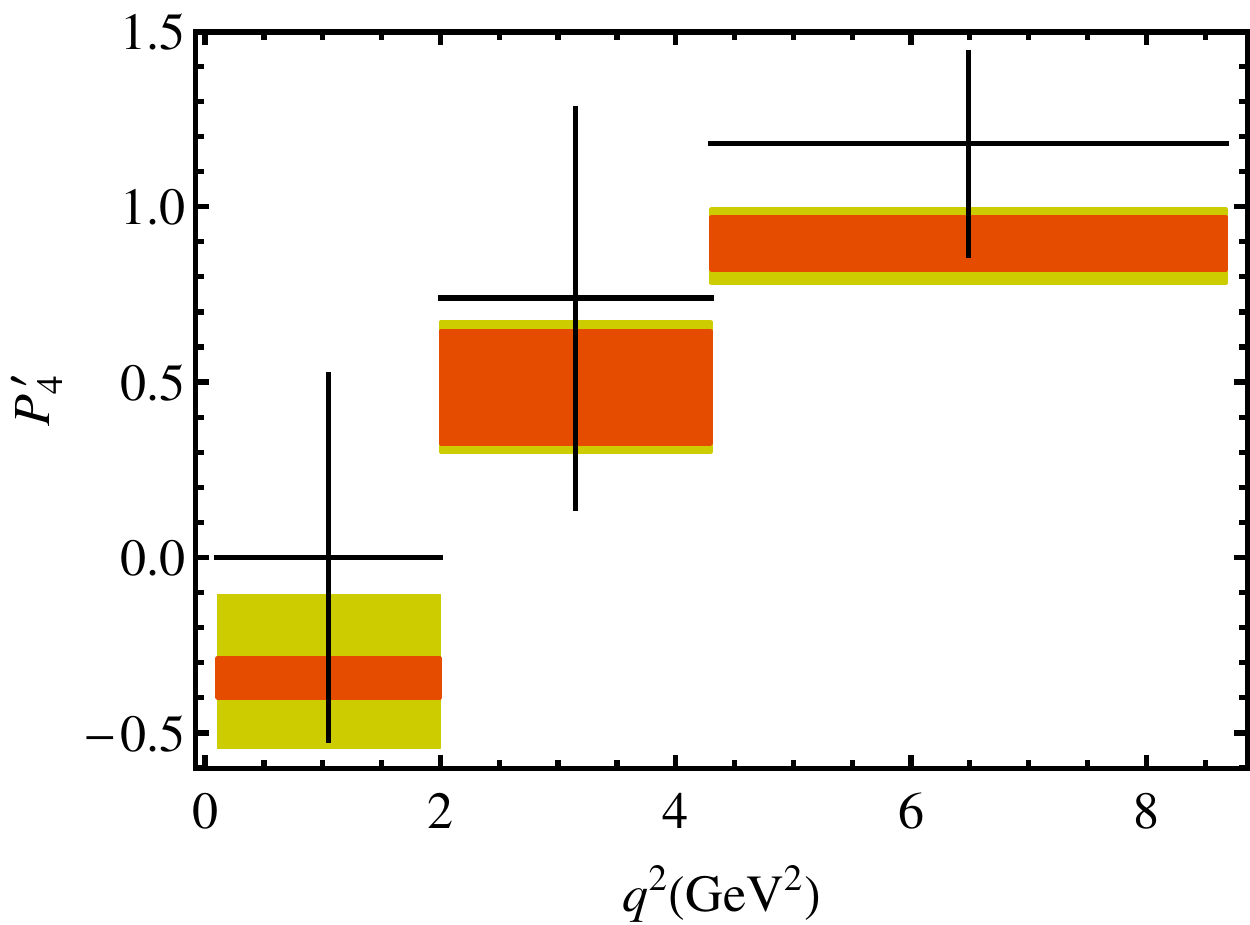}\\[5mm]
\includegraphics[width=7cm,height=4.6cm]{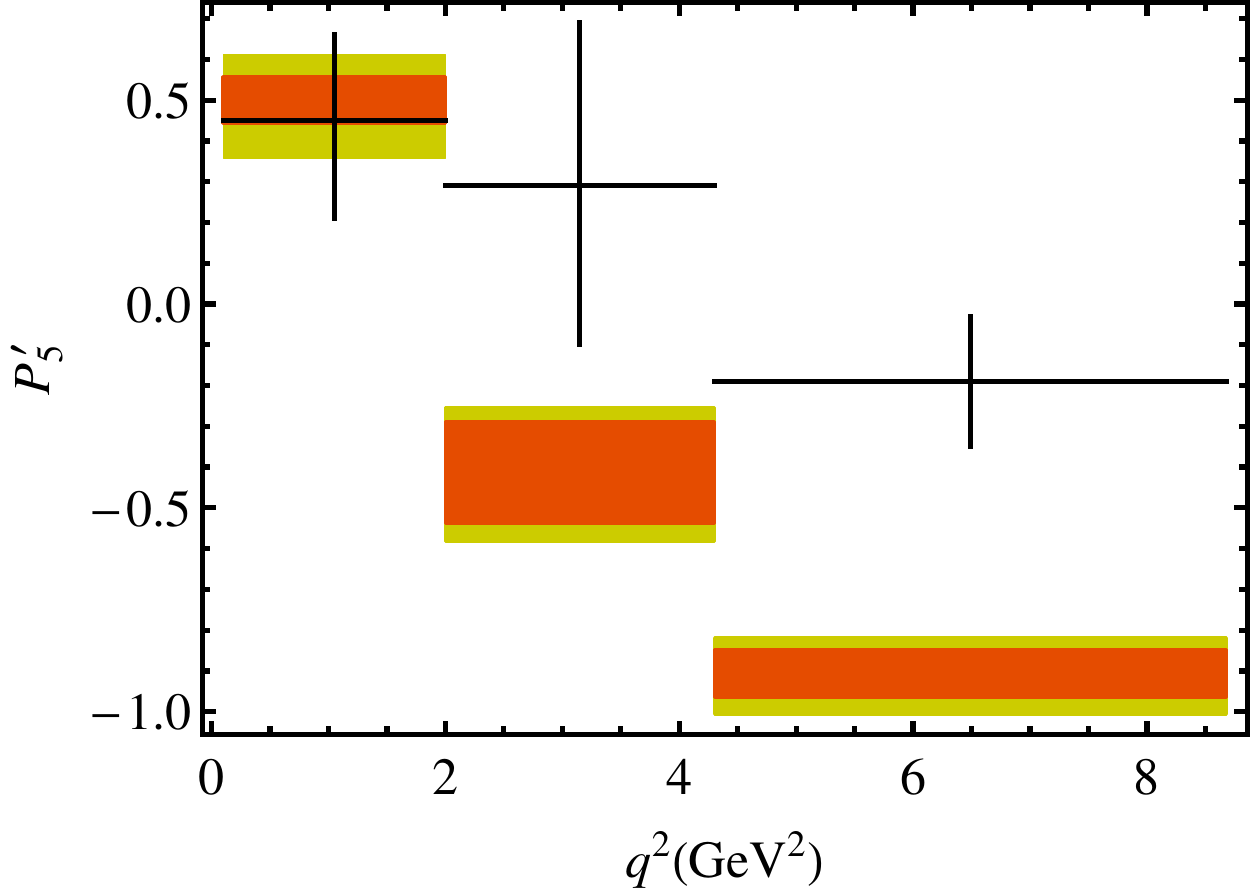}
\caption{SM predictions for the observables $P_1$, $P_2$, $P_4'$, $P_5'$. The bands correspond
to all uncertainties added in quadrature, not including (dark) and including (light) our
 estimate of long-distance charm-loop effects. The data
points correspond to experimental data from LHCb~\cite{Aaij:2013iag,Aaij:2013qta}.
\label{fig:res2}}
\end{figure}

In fig.~\ref{fig:res2} we show our results where the long-distance $c\bar{c}$ correction is displayed as a 
separate band added in quadrature to the combined error from parametric uncertainties, form factors and 
power corrections. These plots constitute our predictions including charm-loop effects. 
We note that the discrepancy between theory and data in the third bin of $P_5^{\prime}$ remains also when
QCD uncertainties from power corrections and from charm loops are included in the theory prediction. 
Results for all the $S_i$- and $P_i^{(\prime)}$-observables and further details can be found 
in ref.~\cite{Descotes-Genon:2014uoa}.

\section{Conclusions}
\label{sec:conclu}
A QCDF-improved calculation based on soft form factors allows for precise predictions of
$B\to K^*\mu^+\mu^-$ observables even in the absence of knowledge on correlations of the
form factor uncertainties. 

In optimised observables large-recoil symmetries enforce a cancellation
of the form factors at LO in $\alpha_s$ and $\Lambda/m_b$ and hence such observables exhibit a reduced 
sensitivity to form factor uncertainties. They are thus sensitive to subleading 
power corrections of order $\Lambda/m_b$ for which this suppression mechanism breaks down as
they break the large-recoil symmtries. We have presented a systematic approach to include factorisable power
corrections into a calculation based on soft form factors. We have further demonstrated that
the impact of factorisable power corrections can be reduced, i.e. the precision of the predictions of
observables can be increased, by a suitable choice of the renormalisation scheme for the
soft form factors. 
 
Finally we have discussed long-distance effects from charm loops based on the partial
calculation of ref.~\cite{Khodjamirian:2010vf} whose results we use as an estimator for the expected 
order of magnitude.

Our complete results for two different sets of LCSR form factors \cite{Ball:2004ye,Khodjamirian:2010vf} 
can be found in ref.~\cite{Descotes-Genon:2014uoa}.

\section*{Acknowledgments}
J.V. is funded by the Deutsche Forschungsgemeinschaft (DFG) within research unit FOR 1873 (QFET).
J.M. and L.H. acknowledge support from FPA2011-25948, SGR2009-00894.




\nocite{*}
\bibliographystyle{elsarticle-num}



\end{document}